\documentclass[11pt]{article}

\usepackage[margin=1in]{geometry}
\usepackage[T1]{fontenc}
\usepackage[utf8]{inputenc}
\usepackage{lmodern}
\usepackage{microtype}
\usepackage{setspace}

\usepackage{authblk}

\usepackage{amsmath,amssymb}
\usepackage{siunitx}
\usepackage{graphicx}
\usepackage{booktabs}
\usepackage{array}
\usepackage{tabularx}
\usepackage{longtable}
\usepackage{adjustbox}
\usepackage{multirow}
\usepackage{makecell}
\usepackage{threeparttable}
\usepackage{float}
\usepackage{placeins}
\usepackage{caption}
\usepackage{subcaption}

\newcolumntype{Y}{>{\raggedright\arraybackslash}X}

\usepackage{enumitem}
\setlist[itemize]{leftmargin=1.5em}
\setlist[enumerate]{leftmargin=1.5em}

\usepackage{xcolor}
\usepackage{xurl}
\usepackage[numbers,sort&compress]{natbib}
\usepackage{hyperref}
\usepackage[nameinlink,noabbrev]{cleveref}

\hypersetup{
    colorlinks=true,
    linkcolor=blue!50!black,
    citecolor=blue!50!black,
    urlcolor=blue!50!black,
    pdftitle={Fourier Morphology Diagnostics of Delta Cephei MESA-RSP Light Curves},
    pdfauthor={Zuhoor Elahi, Christopher Sirola, Wafa Gull}
}

\graphicspath{{./}}

\newcommand{\safeincludegraphics}[2][]{%
    \IfFileExists{#2}{%
        \includegraphics[#1]{#2}%
    }{%
        \fbox{\begin{minipage}[c][0.28\textheight][c]{0.88\textwidth}
        \centering \small Missing figure file:\\[0.5em]\texttt{\detokenize{#2}}
        \end{minipage}}%
    }%
}

\newcommand{\dcep}{\ensuremath{\delta} Cephei}

\newcommand{\Lsun}{\ensuremath{L_{\odot}}}
\newcommand{\Teff}{\ensuremath{T_{\rm eff}}}

\newcommand{\logg}{\ensuremath{\log g}}

\newcommand{\Mbol}{\ensuremath{M_{\rm bol}}}
\newcommand{\BC}{\ensuremath{\mathrm{BC}}}
\newcommand{\AV}{\ensuremath{A_V}}
\newcommand{\feh}{\ensuremath{[{\rm Fe/H}]}}
\newcommand{\MESA}{\textsc{MESA}}
\newcommand{\RSP}{\textsc{RSP}}
\newcommand{\MESARSP}{\MESA-\RSP}
\newcommand{\MIST}{\textsc{MIST}}

\newcommand{\code}[1]{\texttt{\detokenize{#1}}}

\newcommand{\RSPalfam}{\ifmmode\mathrm{RSP\_alfam}\else\texttt{RSP\_alfam}\fi}
\newcommand{\RSPalfat}{\ifmmode\mathrm{RSP\_alfat}\else\texttt{RSP\_alfat}\fi}
\newcommand{\RSPgammar}{\ifmmode\mathrm{RSP\_gammar}\else\texttt{RSP\_gammar}\fi}
\newcommand{\alfam}{\RSPalfam}
\newcommand{\alfat}{\RSPalfat}
\newcommand{\gammar}{\RSPgammar}

\title{Fourier Morphology Diagnostics of Delta Cephei MESA-RSP Light Curves}

\author[1,2,*]{Zuhoor Elahi}
\author[1]{Christopher Sirola}
\author[1]{Wafa Gull}

\affil[1]{Department of Physics and Astronomy, University of Southern Mississippi, Hattiesburg, MS, USA}
\affil[2]{Department of Physics, University of Karachi, Karachi, Pakistan}
\affil[*]{Corresponding author: zuhoor.elahi@usm.edu}

\date{}

\begin{document}
\maketitle

\begin{abstract}
Fourier morphology provides a compact test of whether a nonlinear Cepheid model reproduces the observed harmonic content and asymmetric light-curve shape, not just the pulsation period. We apply Fourier and time-domain morphology diagnostics to observed and synthetic \dcep{} light curves. The observed Johnson \(V\)-band light curve is represented by a Fourier template and compared with MIST bolometric-correction \(V\)-band light curves generated from the calibrated \MESA-\RSP{} model sequence. This study is framed as a target-specific morphology assessment rather than as a new Fourier-decomposition method. The observed template has \(\Delta V=0.8390~{\rm mag}\), \(R_{21}=0.3401\), \(R_{31}=0.1943\), rise fraction \(0.2775\), and asymmetry index \(0.4450\). The final accepted model, with \(\alfam=0.400\) and \(\alfat=0.095\), gives \(\Delta V_{\rm syn}=0.0367~{\rm mag}\), \(R_{21}=0.0185\), \(R_{31}=0.0028\), rise fraction \(0.4860\), and asymmetry index \(0.0280\). The accepted calibration sequence therefore improves the synthetic amplitude only modestly and leaves the synthetic curve much too sinusoidal. The result shows that the accepted sequence is period-calibrated and morphology-tested, but it is not yet an observed-amplitude solution for \dcep{}.
\end{abstract}


\section{Introduction}
\label{sec:introduction}

Classical Cepheids are radially pulsating evolved stars whose pulsation periods are connected to their luminosities. They are therefore central to the extragalactic distance scale and also provide important tests of stellar evolution and nonlinear pulsation physics.  \dcep{} is the prototype of the class and is an especially useful benchmark because its period, brightness variation, and characteristic asymmetric light curve are well observed.

The distance-scale motivation for Cepheid work follows from the period--luminosity relation and its use in the modern extragalactic distance ladder \citep{leavitt1912,freedmanmadore1991,freedman2001,pietrzynski2019,riess2022}. The modeling motivation is complementary: nonlinear pulsation calculations and time-dependent convection treatments are needed to explain not just the period but also amplitudes, phase structure, and morphology \citep{cox1980,stellingwerf1982,smolecmoskalik2008,paxton2019,jermyn2023}. This work uses those established tools as context for a focused Delta-Cephei morphology assessment.

A successful Cepheid model must reproduce more than the fundamental pulsation period.  In a nonlinear model, the observed-band light curve also depends on the amplitude of the radius and temperature variations, the phase relation among radius, luminosity, and effective temperature, and the treatment of time-dependent convection and damping.  Therefore, a model may be period-calibrated while still failing to reproduce the observed amplitude or the observed non-sinusoidal shape.

Fourier decomposition provides a compact way to quantify Cepheid light-curve morphology.  The amplitude ratios \(R_{21}\) and \(R_{31}\) measure the relative strength of higher harmonics, while the phase combinations \(\phi_{21}\) and \(\phi_{31}\) describe the phase structure of the curve.  Additional time-domain measures such as rise fraction and asymmetry index directly quantify the rapid rise and slower decline that are characteristic of classical Cepheid light curves.

This paper focuses on Fourier morphology diagnostics and parameter calibration with the Modules for Experiments in Stellar Astrophysics (MESA) Radial Stellar Pulsation (RSP) module, hereafter \MESARSP{}, for \dcep.  The period calibration and basic \MESARSP{} setup are described in the companion period-calibration analysis \citep{elahi2026nonlinear}, while the synthetic observed-band light-curve construction with MESA Isochrones and Stellar Tracks (MIST) bolometric-correction tables, hereafter MIST-BC tables, is described in the companion synthetic-photometry analysis \citep{elahi2026synthetic}.  Here, the scientific question is narrower: can the accepted \MESARSP{} models reproduce the observed Johnson \(V\)-band morphology of \dcep{} after synthetic photometry is applied?

The main result is that the final accepted model improves the synthetic amplitude and morphology only modestly.  Here \(\Delta V\) denotes the peak-to-peak Johnson/Bessell \(V\)-band magnitude amplitude.  The best accepted model in this calibration stage has \(\alfam=0.400\) and \(\alfat=0.095\), but still produces only \(\Delta V_{\rm syn}=0.0367~{\rm mag}\), compared with the observed \(\Delta V_{\rm obs}=0.8390~{\rm mag}\).  Its Fourier ratios and asymmetry index are also much smaller than the observed values.  Therefore, the final model is morphology-tested and improved, but it is not an observed-amplitude solution.

\subsection*{Scope and relation to existing Fourier work}
Fourier decomposition of Cepheid light curves is a standard diagnostic, with a long history in the interpretation of Cepheid light-curve structure \citep{simonlee1981,antonelloporetti1986,kanburngeow2004}. Multiwavelength comparisons between theoretical and observed Cepheid light curves, including Fourier-parameter comparisons and phase-dependent relations, have been developed extensively \citep[e.g.,][]{bhardwaj2017,kurbah2023,hocde2024,deka2025}. The contribution of the present work is a focused, target-specific assessment of the accepted Delta-Cephei \MESA-\RSP{} sequence after synthetic photometry has been applied. In particular, the paper quantifies how the residual amplitude mismatch appears in the Fourier ratios, rise fraction, and asymmetry index.

\section{Observational and Synthetic Light-Curve Data}
\label{sec:data}

\subsection{Observed Johnson \texorpdfstring{\(V\)}{V}-band light curve}
\label{subsec:observed_data}

The observational benchmark used in this paper is the cleaned American Association of Variable Star Observers (AAVSO) Johnson \(V\)-band light curve of \dcep{} \citep{aavso}.  The data were filtered to retain Johnson \(V\)-band measurements, cleaned using magnitude and uncertainty cuts, phase-folded using the adopted observed period,
\begin{equation}
    P_{\rm obs}=5.366531~{\rm d},
    \label{eq:pobs}
\end{equation}
and represented by a smoothed Fourier template.  The resulting fitted peak-to-peak amplitude is
\begin{equation}
    \Delta V_{\rm obs}=0.8390~{\rm mag}.
    \label{eq:delta_v_obs}
\end{equation}
The observed Fourier fit gives
\begin{equation}
    R_{21,\rm obs}=0.3401,\qquad
    R_{31,\rm obs}=0.1943,
    \label{eq:observed_fourier_ratios}
\end{equation}
with
\begin{equation}
    \phi_{21,\rm obs}=4.4477~{\rm rad},\qquad
    \phi_{31,\rm obs}=2.3642~{\rm rad}.
    \label{eq:observed_fourier_phases}
\end{equation}
The rise fraction and asymmetry index are
\begin{equation}
    f_{\rm rise,obs}=0.2775,\qquad
    A_{\rm asym,obs}=0.4450.
    \label{eq:observed_asymmetry}
\end{equation}

\begin{figure}[htbp]
    \centering
    \includegraphics[width=0.95\textwidth]{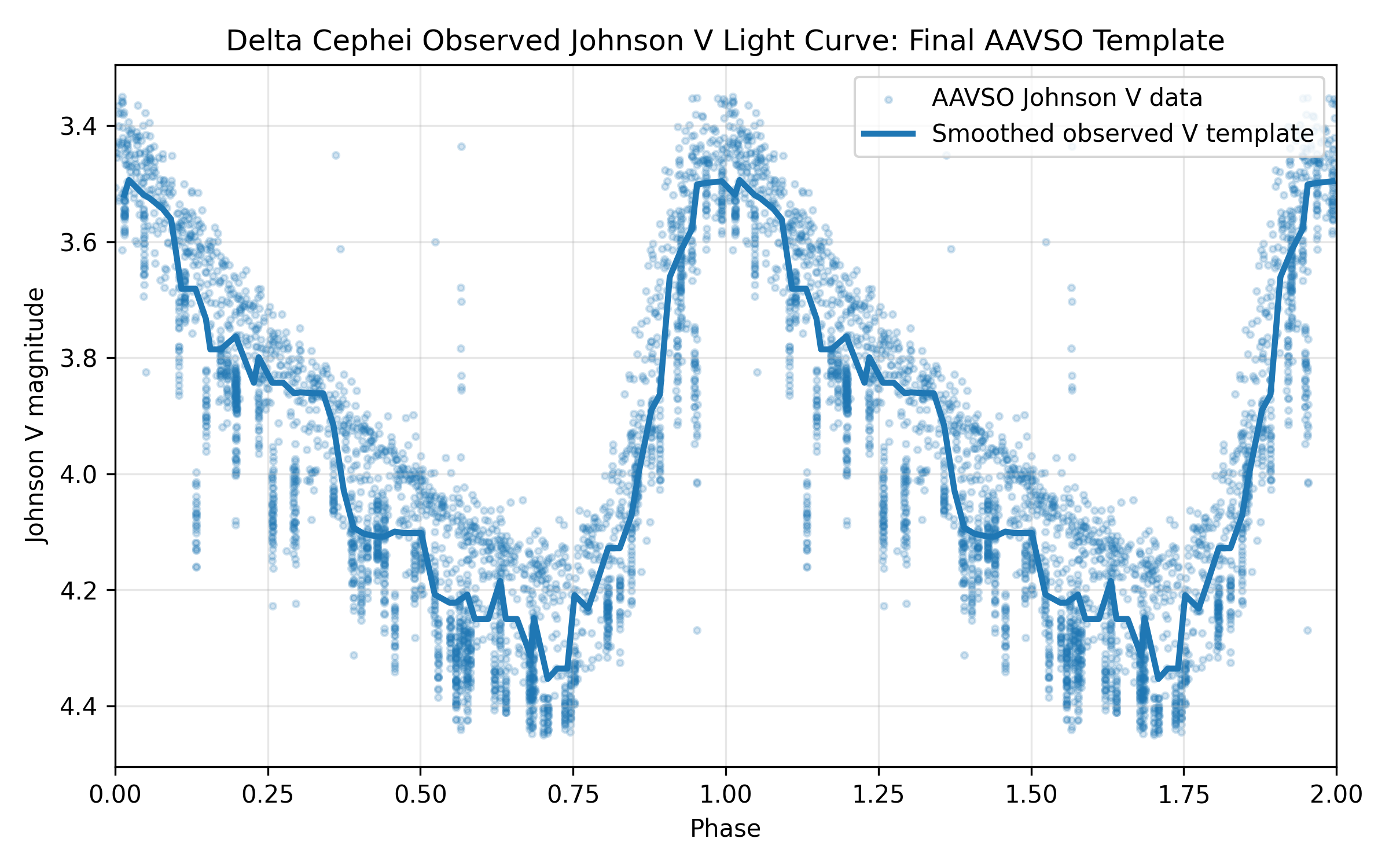}
    \caption{Observed Johnson \(V\)-band light curve of \dcep{} from the cleaned AAVSO data.  The points show the selected observations, while the solid curve shows the robust smoothed phase template used as the observational morphology benchmark.}
    \label{fig:observed_v_template}
\end{figure}

\begin{figure}[htbp]
    \centering
    \includegraphics[width=0.95\textwidth]{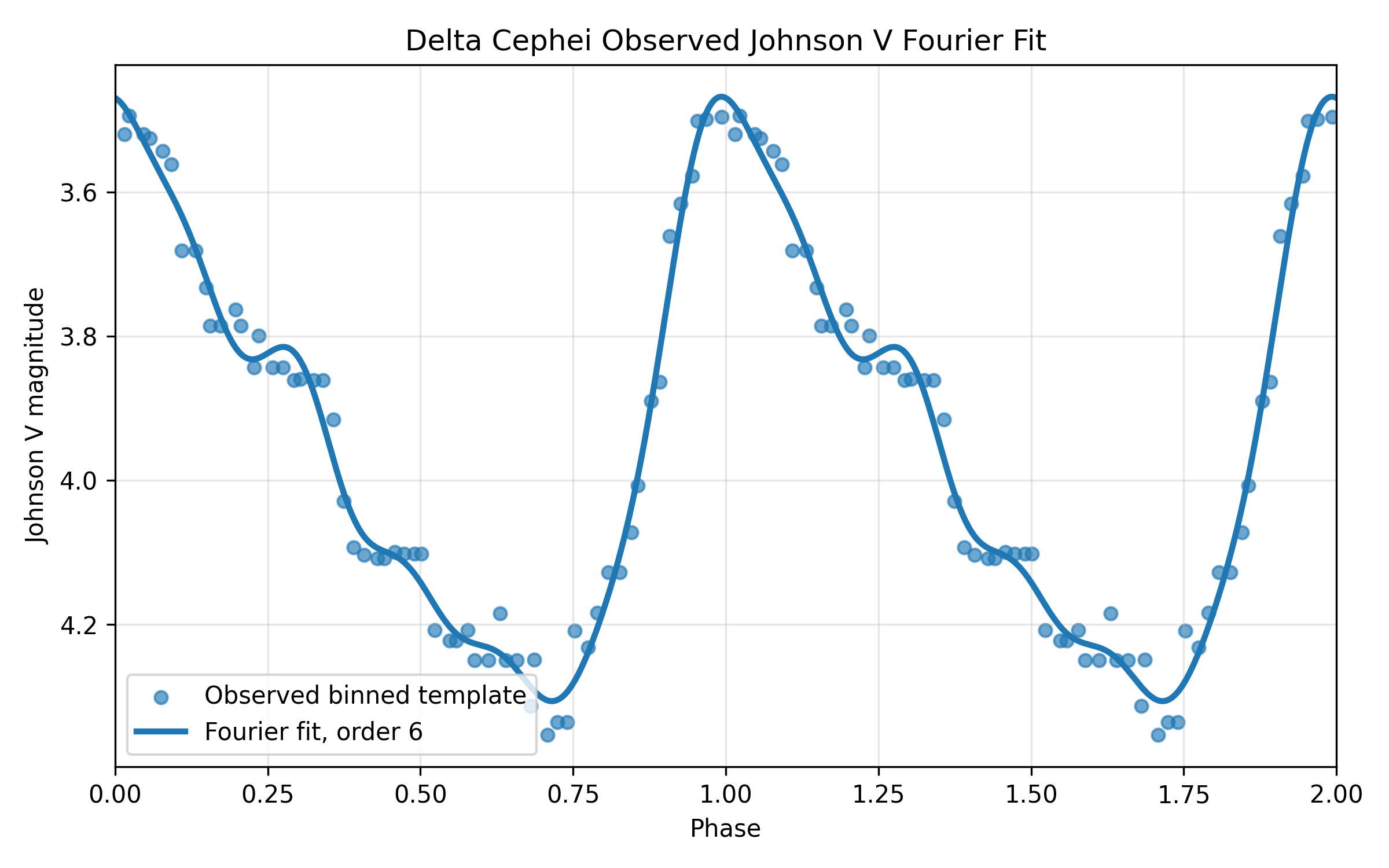}
    \caption{Fourier fit to the observed Johnson \(V\)-band light curve of \dcep.  The fit quantifies the observed amplitude, harmonic content, and asymmetric Cepheid morphology.}
    \label{fig:observed_v_fourier_fit}
\end{figure}

\FloatBarrier

\subsection{Synthetic MESA-RSP light curves}
\label{subsec:synthetic_data}

The synthetic light curves were generated from the phase-dependent \MESARSP{} luminosity \(L(\phi)\), effective temperature \(\Teff(\phi)\), and radius \(R(\phi)\),
\begin{equation}
    L(\phi),\qquad \Teff(\phi),\qquad R(\phi),
\end{equation}
where \(\phi\) is pulsation phase.  The bolometric magnitude \(\Mbol\) was computed from the luminosity as
\begin{equation}
    M_{\rm bol}(\phi)
    =M_{{\rm bol},\odot}
    -2.5\log_{10}\left[\frac{L(\phi)}{\Lsun}\right].
    \label{eq:mbol}
\end{equation}
The synthetic bandpass magnitude was then computed using \MIST{} bolometric corrections \citep{choi2016,dotter2016}.  In the following expression, \(\BC_{\lambda}\) is the bolometric correction in bandpass \(\lambda\), \(\logg\) is the surface gravity, \(\feh\) is the logarithmic iron abundance relative to the solar value, and \(\AV\) is the visual extinction:
\begin{equation}
    M_{\lambda}(\phi)=M_{\rm bol}(\phi)-\BC_{\lambda}\left[\Teff(\phi),\log g(\phi),\feh,\AV\right].
    \label{eq:mist_bc_mag}
\end{equation}
For this paper, the primary synthetic comparison is the \MIST{} bolometric-correction Bessell \(V\)-band curve \citep{bessell1990,bessellmurphy2012}, denoted here as synthetic Johnson/Bessell \(V\).

\subsection{Accepted and diagnostic calibration models}
\label{subsec:model_sequence}

The accepted morphology-calibration sequence is listed in Table~\ref{tab:accepted_sequence}.  The \(\alfam\) parameter is the \MESARSP{} eddy-viscous damping control parameter used to adjust the nonlinear pulsation-amplitude response, while \(\alfat\) is the \MESARSP{} turbulent-transport/control parameter varied in the final morphology-tuning sequence.  The sequence begins with the enhanced-amplitude \(\alfam=0.425\) case, proceeds to \(\alfam=0.400\), and then refines \(\alfat\).  The final accepted model is the \(\alfam=0.400\), \(\alfat=0.095\) case, which satisfied the adopted quality-control criteria.

\begin{table}[htbp]
\centering
\caption{Accepted MESA-RSP morphology-calibration sequence.  The amplitude values refer to the synthetic \MIST{} bolometric-correction \(V\)-band peak-to-peak amplitude.}
\label{tab:accepted_sequence}
\begin{tabular}{lcccc}
\toprule
Model & \(\alfam\) & \(\alfat\) & \(\Delta V_{\rm syn}\) [mag] & Model role \\
\midrule
M1 & 0.425 & default & 0.0302 & accepted sequence model \\
M2 & 0.400 & default & 0.0344 & accepted sequence model \\
M3 & 0.400 & 0.075 & 0.0352 & accepted sequence model \\
M4 & 0.400 & 0.095 & 0.0367 & final accepted model \\
\bottomrule
\end{tabular}
\end{table}

Two higher-\(\alfat\) cases were examined but were not accepted as final quantitative models. They are retained only as diagnostic cases because they did not satisfy the adopted quality-control criteria.

\begin{table}[htbp]
\centering
\caption{Diagnostic or rejected morphology-calibration models.}
\label{tab:diagnostic_models}
\begin{tabular}{lccc}
\toprule
Model & \(\alfam\) & \(\alfat\) & Reason for non-acceptance \\
\midrule
D1 & 0.400 & 0.0975 & warning or numerical-stability issue \\
D2 & 0.400 & 0.1000 & warning or numerical-stability issue \\
\bottomrule
\end{tabular}
\end{table}

\section{Fourier and Morphology Diagnostics}
\label{sec:methods}

\subsection{Fourier representation}
\label{subsec:fourier_representation}

Each phased magnitude light curve was represented using a truncated Fourier series,
\begin{equation}
    m(\phi)=A_0+\sum_{k=1}^{N}A_k\cos\left(2\pi k\phi+\varphi_k\right),
    \label{eq:fourier_series}
\end{equation}
where \(m(\phi)\) is the magnitude at pulsation phase \(\phi\), \(A_k\) is the amplitude of the \(k\)-th harmonic, and \(\varphi_k\) is its phase.  The same diagnostic definitions were applied to the observed Fourier template and the synthetic \MIST{} bolometric-correction \(V\)-band curves.

\subsection{Fourier amplitude ratios}
\label{subsec:amplitude_ratios}

The Fourier amplitude ratios are defined as
\begin{equation}
    R_{k1}=\frac{A_k}{A_1}.
    \label{eq:fourier_ratio_general}
\end{equation}
The two primary ratios used here are
\begin{equation}
    R_{21}=\frac{A_2}{A_1},\qquad
    R_{31}=\frac{A_3}{A_1}.
    \label{eq:r21_r31}
\end{equation}
Large values of \(R_{21}\) and \(R_{31}\) indicate strong higher-harmonic content and therefore a more non-sinusoidal light-curve shape.  A nearly sinusoidal model has small \(R_{21}\) and \(R_{31}\).

\subsection{Fourier phase combinations}
\label{subsec:phase_combinations}

The Fourier phase combinations are defined as
\begin{equation}
    \phi_{k1}=\varphi_k-k\varphi_1,
    \label{eq:phase_combination_general}
\end{equation}
with
\begin{equation}
    \phi_{21}=\varphi_2-2\varphi_1,
    \qquad
    \phi_{31}=\varphi_3-3\varphi_1.
    \label{eq:phi21_phi31}
\end{equation}
These phase combinations provide information about the phase structure and asymmetry of the light curve beyond amplitude alone.

\subsection{Rise fraction and asymmetry index}
\label{subsec:rise_asymmetry}

Because magnitudes decrease as brightness increases, maximum brightness corresponds to minimum magnitude.  The rise fraction is measured as the fraction of the pulsation cycle required for the star to rise from minimum brightness to maximum brightness.  In the phased magnitude curve this corresponds to the phase interval from maximum magnitude to minimum magnitude, wrapped over one cycle if necessary.

The decline fraction is
\begin{equation}
    f_{\rm decline}=1-f_{\rm rise}.
    \label{eq:decline_fraction}
\end{equation}
The asymmetry index used in this paper is
\begin{equation}
    A_{\rm asym}
    =\frac{|f_{\rm rise}-f_{\rm decline}|}{f_{\rm rise}+f_{\rm decline}}
    =|2f_{\rm rise}-1|.
    \label{eq:asymmetry_index}
\end{equation}
A symmetric sinusoidal curve has \(f_{\rm rise}\simeq 0.5\) and \(A_{\rm asym}\simeq 0\), while a strongly asymmetric Cepheid curve has a smaller rise fraction and a larger asymmetry index.

\section{Results}
\label{sec:results}

\subsection{Observed Fourier morphology of Delta Cephei}
\label{subsec:observed_morphology}

The observed Johnson \(V\)-band Fourier template is strongly non-sinusoidal.  Table~\ref{tab:observed_morphology} summarizes the measured observed morphology diagnostics.  The values \(R_{21}=0.3401\) and \(R_{31}=0.1943\) show that the second and third harmonics contribute significantly to the light-curve shape.  The rise fraction of \(0.2775\) indicates a rapid rise to maximum brightness followed by a slower decline.

\begin{table}[htbp]
\centering
\caption{Observed Johnson \(V\)-band morphology diagnostics for \dcep.}
\label{tab:observed_morphology}
\begin{tabular}{lc}
\toprule
Quantity & Observed value \\
\midrule
\(\Delta V\) & \(0.8390~{\rm mag}\) \\
\(R_{21}\) & 0.3401 \\
\(R_{31}\) & 0.1943 \\
\(\phi_{21}\) & 4.4477 rad \\
\(\phi_{31}\) & 2.3642 rad \\
Rise fraction & 0.2775 \\
Decline fraction & 0.7225 \\
Asymmetry index & 0.4450 \\
\bottomrule
\end{tabular}
\end{table}

\subsection{Final accepted model morphology}
\label{subsec:final_model_morphology}

The final accepted model has
\begin{equation}
    \alfam=0.400,\qquad \alfat=0.095,
    \label{eq:final_model_params}
\end{equation}
and was selected because it gave the largest accepted synthetic \(V\)-band amplitude in the model sequence.  Its measured morphology diagnostics are
\begin{equation}
    \Delta V_{\rm syn}=0.0367~{\rm mag},\qquad
    R_{21,\rm syn}=0.0185,
    \qquad
    R_{31,\rm syn}=0.0028,
    \label{eq:final_model_fourier_ratios}
\end{equation}
with
\begin{equation}
    \phi_{21,\rm syn}=4.3190~{\rm rad},\qquad
    \phi_{31,\rm syn}=5.1895~{\rm rad},
    \label{eq:final_model_fourier_phases}
\end{equation}
and
\begin{equation}
    f_{\rm rise,syn}=0.4860,
    \qquad
    A_{\rm asym,syn}=0.0280.
    \label{eq:final_model_asymmetry}
\end{equation}
These values show that the final synthetic curve remains close to sinusoidal.  In particular, the rise fraction is close to \(0.5\), and the asymmetry index is close to zero.

\begin{table}[htbp]
\centering
\caption{Observed Johnson \(V\)-band morphology compared with the final accepted \MESA-\RSP{} \MIST-BC synthetic \(V\)-band model.}
\label{tab:observed_vs_final}
\begin{tabular}{lccc}
\toprule
Quantity & Observed & Final model & Model/Observed \\
\midrule
\(\Delta V\) [mag] & 0.8390 & 0.0367 & 0.0437 \\
\(R_{21}\) & 0.3401 & 0.0185 & 0.0544 \\
\(R_{31}\) & 0.1943 & 0.0028 & 0.0144 \\
\(\phi_{21}\) [rad] & 4.4477 & 4.3190 & 0.9711 \\
\(\phi_{31}\) [rad] & 2.3642 & 5.1895 & 2.1950 \\
Rise fraction & 0.2775 & 0.4860 & 1.7514 \\
Decline fraction & 0.7225 & 0.5140 & 0.7114 \\
Asymmetry index & 0.4450 & 0.0280 & 0.0629 \\
\bottomrule
\end{tabular}
\end{table}

\begin{figure}[htbp]
    \centering
    \includegraphics[width=0.95\textwidth]{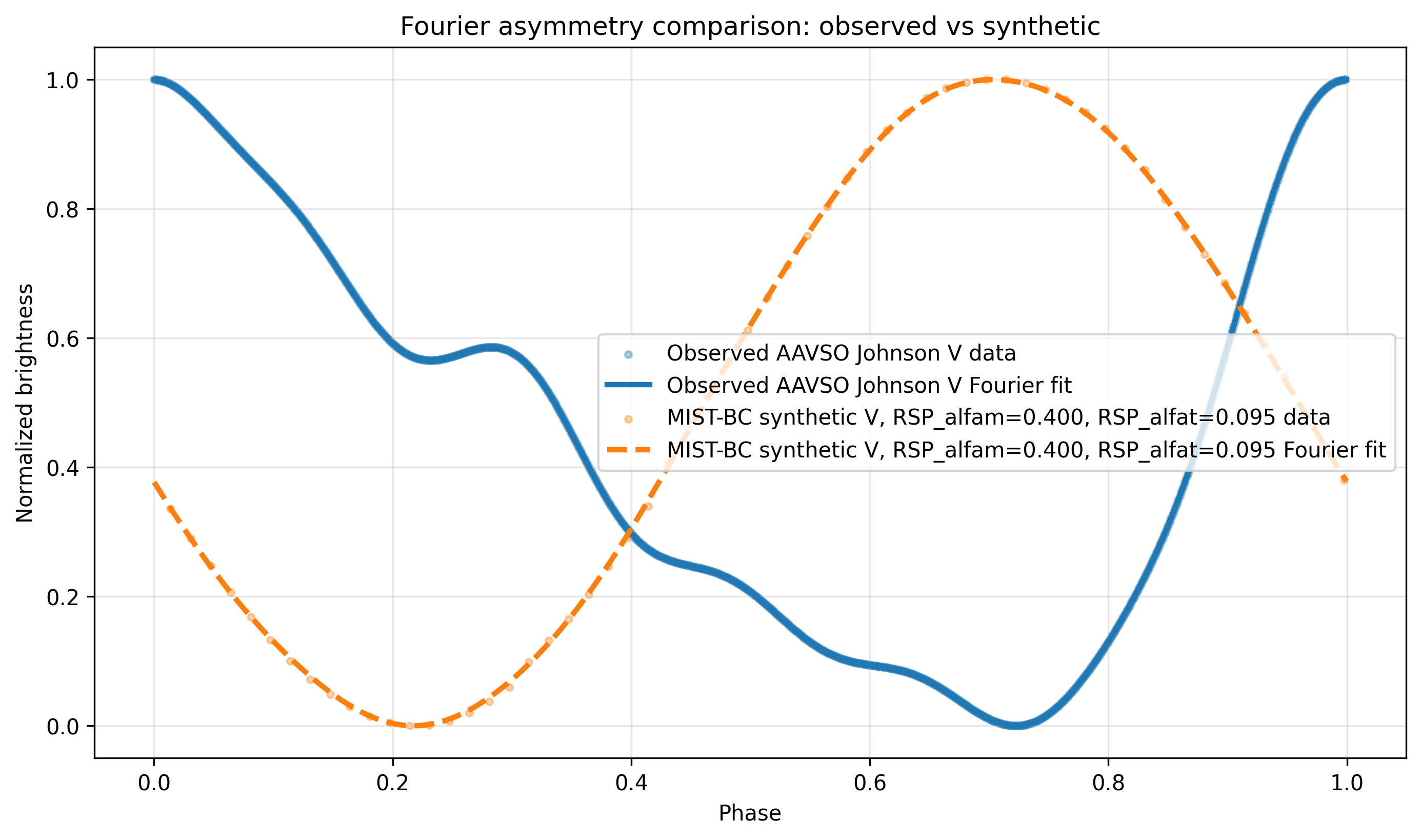}
    \caption{Fourier morphology comparison between the observed AAVSO Johnson \(V\)-band template and the final accepted synthetic \MIST-BC \(V\)-band model with \(\alfam=0.400\) and \(\alfat=0.095\).  The observed curve has much stronger harmonic content and asymmetry, while the synthetic curve remains much smoother and more nearly sinusoidal.}
    \label{fig:fourier_asymmetry_final}
\end{figure}

\FloatBarrier

\subsection{Harmonic-content progression across the accepted sequence}
\label{subsec:harmonic_content_progression}

The companion synthetic-photometry paper presents the accepted-model amplitude progression as an observed-band amplitude result.  To avoid duplicating that amplitude-only figure here, the present paper uses the accepted sequence to examine the Fourier morphology response instead.  The key question is whether the calibration sequence increases the second-harmonic content enough to approach the observed Johnson \(V\)-band morphology.

The observed template has \(R_{21,\rm obs}=0.3401\), while the final accepted model has \(R_{21,\rm syn}=0.0185\).  Thus the final model reaches only
\begin{equation}
    \frac{R_{21,\rm syn}}{R_{21,\rm obs}}
    =\frac{0.0185}{0.3401}
    \simeq 0.0544,
    \label{eq:r21_fraction}
\end{equation}
or about \(5.4\%\) of the observed second-harmonic amplitude ratio.  This diagnostic is more appropriate for the present morphology paper than a repeated \(\Delta V\)-only progression plot, because it directly measures how far the accepted synthetic curves remain from the observed non-sinusoidal Cepheid shape.

\begin{figure}[htbp]
    \centering
    \includegraphics[width=0.88\textwidth]{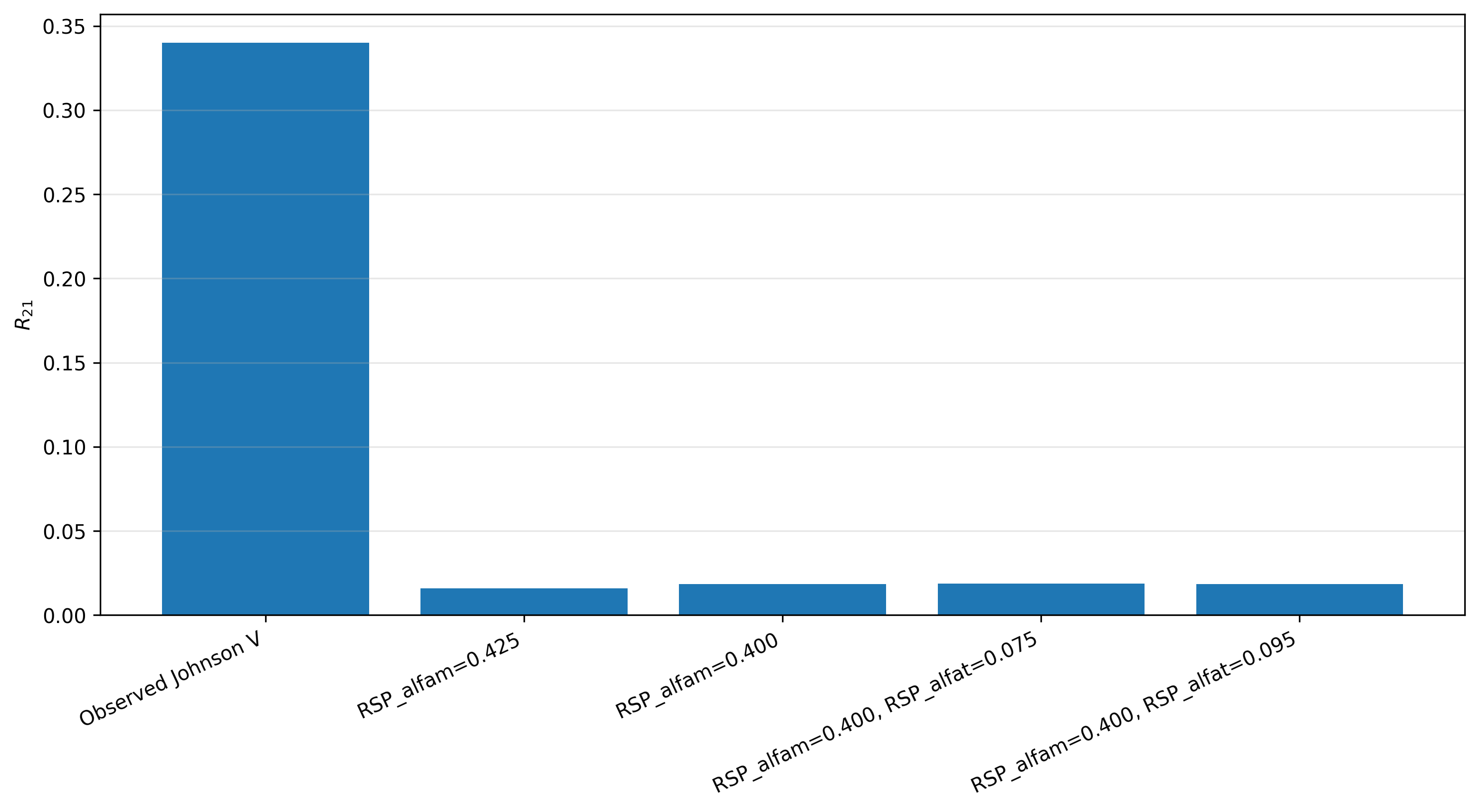}
    \caption{Progression of the Fourier amplitude ratio \(R_{21}\) across the accepted morphology-calibration sequence.  The observed Johnson \(V\)-band template has much stronger second-harmonic content than any accepted synthetic model.  Although the accepted \MESA-\RSP{} sequence changes \(R_{21}\), the final accepted model remains far too sinusoidal compared with \dcep.}
    \label{fig:final_R21_progression}
\end{figure}

\begin{figure}[htbp]
    \centering
    \includegraphics[width=0.95\textwidth]{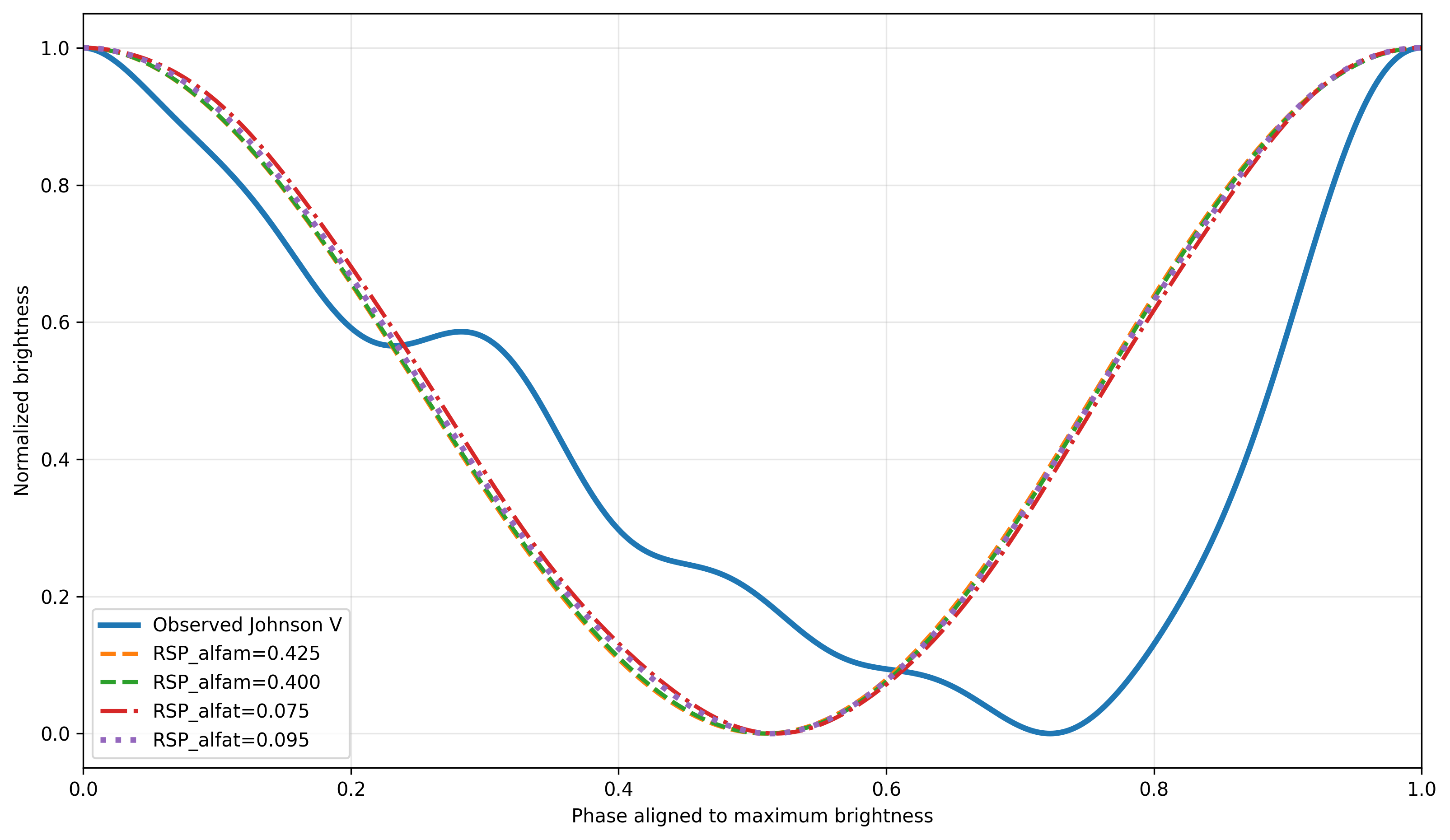}
    \caption{Phase-aligned morphology progression for the accepted model sequence.  The curves are normalized and aligned at maximum brightness, so the figure compares shape rather than absolute amplitude.  The accepted synthetic curves remain much smoother and more nearly sinusoidal than the observed Johnson \(V\)-band curve.}
    \label{fig:final_phase_aligned_progression}
\end{figure}

\FloatBarrier

\subsection{Diagnostic high-growth cases}
\label{subsec:diagnostic_high_growth}

The higher-\(\alfat\) cases, \(\alfat=0.0975\) and \(\alfat=0.100\), were not used as accepted final comparisons.  Although such models can be useful for understanding local sensitivity to \RSP{} parameter changes, they were treated as diagnostic only because of warnings or numerical-stability issues.  This distinction is important: the purpose of the calibration sequence is not simply to maximize amplitude, but to identify models that improve the synthetic light curve while preserving a stable, reproducible run.

\section{Discussion}
\label{sec:discussion}

\subsection{Period matching is not enough}
\label{subsec:period_not_enough}

The final accepted model can be considered successful in a limited period-calibration sense, but not in a morphology-calibration sense.  The large discrepancy in \(\Delta V\), \(R_{21}\), \(R_{31}\), and asymmetry index shows that reproducing the correct global pulsation period does not guarantee that the model has the correct nonlinear light-curve structure.

The amplitude mismatch is especially severe: the final accepted synthetic \(V\)-band amplitude is only about \(4.4\%\) of the observed amplitude.  The morphology mismatch is also severe: \(R_{21}\) is only about \(5.4\%\) of the observed value, \(R_{31}\) is only about \(1.4\%\) of the observed value, and the synthetic asymmetry index is only about \(6.3\%\) of the observed value.  The synthetic light curve is therefore dominated by the fundamental sinusoidal component and lacks the higher-harmonic structure required to reproduce the observed Cepheid morphology.

\subsection{Interpretation of weak harmonic content}
\label{subsec:weak_harmonics}

The small values of \(R_{21}\) and \(R_{31}\) indicate that the final synthetic light curve contains very little higher-harmonic power relative to the fundamental term.  This explains why the synthetic curve remains smooth and sinusoidal even after the amplitude-control sequence.  In observational Cepheids, the nonzero second and third harmonics encode the rapid rise and slower decline of the brightness variation.  In the present model, that harmonic structure is strongly underproduced.

\subsection{Role of RSP parameter calibration}
\label{subsec:rsp_parameters}

The accepted calibration sequence demonstrates that the selected \RSP{} damping and transport parameters affect the synthetic amplitude and morphology.  Reducing \(\alfam\) from 0.425 to 0.400 increases the synthetic amplitude from \(0.0302\) to \(0.0344~{\rm mag}\), and the subsequent \(\alfat=0.095\) refinement gives \(0.0367~{\rm mag}\).  However, the response remains far too small to match the observed Johnson \(V\)-band amplitude.

The resulting interpretation is that the explored numerically stable parameter range is insufficient.  A broader stable \RSP{} parameter grid is required before claiming an observed-amplitude solution and would include coupled changes in \(\texttt{RSP\_alfa}\), \(\texttt{RSP\_alfac}\), \(\texttt{RSP\_alfas}\), \(\texttt{RSP\_alfad}\), \(\texttt{RSP\_alfap}\), \(\alfam\), \(\alfat\), and the \MESARSP{} \(\gammar\) parameter, while requiring period agreement, late-cycle stability, controlled surface velocities, and absence of severe energy-error warnings.

\subsection{Why the mismatch is not primarily a bolometric-correction problem}
\label{subsec:not_bc_problem}

The use of \MIST{} bolometric corrections is essential because it converts the physical \MESA-\RSP{} outputs into filter-dependent synthetic magnitudes.  However, the final comparison shows that applying bolometric corrections does not remove the amplitude or morphology discrepancy.  The remaining mismatch is therefore primarily a nonlinear pulsation-amplitude and morphology-calibration problem rather than only a photometric-transformation problem.

\subsection{Limitations}
\label{subsec:limitations}

Several limitations follow from the present analysis.  First, the final accepted model is not a complete observed-amplitude Cepheid solution.  Opacity-table sensitivity is not reanalyzed here, but is treated in a separate fixed-model \MESA-\RSP{} opacity comparison \citep{elahi2026opacity}.  Second, the higher-amplitude \(\alfat=0.0975\) and \(\alfat=0.100\) cases are diagnostic only because they did not satisfy the adopted quality-control criteria.  Third, the observational benchmark in this paper is the processed AAVSO Johnson \(V\)-band light curve.  Other filters, including Johnson \(I\), Gaia bands \citep{jordi2010}, and Transiting Exoplanet Survey Satellite (TESS) photometry, require the same cleaning, phase-folding, Fourier-fitting, and comparison procedure before being used for multi-band constraints.

\section{Conclusions}
\label{sec:conclusions}

This paper used Fourier morphology diagnostics to test whether \MESA-\RSP{} synthetic \(V\)-band light curves reproduce the observed Johnson \(V\)-band morphology of \dcep.  The main conclusions are:

\begin{enumerate}[leftmargin=2em]
    \item The observed Johnson \(V\)-band light curve has a large amplitude and strong non-sinusoidal morphology, with \(\Delta V_{\rm obs}=0.8390~{\rm mag}\), \(R_{21}=0.3401\), \(R_{31}=0.1943\), \(f_{\rm rise}=0.2775\), and \(A_{\rm asym}=0.4450\).

    \item The final accepted model, with \(\alfam=0.400\) and \(\alfat=0.095\), gives \(\Delta V_{\rm syn}=0.0367~{\rm mag}\), \(R_{21}=0.0185\), \(R_{31}=0.0028\), \(f_{\rm rise}=0.4860\), and \(A_{\rm asym}=0.0280\).

    \item Across the accepted morphology-calibration sequence, the synthetic light curves show only weak improvement in harmonic content.  In particular, the \(R_{21}\) progression remains far below the observed value, indicating that the accepted models do not develop the second-harmonic structure needed to reproduce the observed Cepheid shape.

    \item The very small \(R_{21}\), \(R_{31}\), and asymmetry index show that the final synthetic light curve remains too sinusoidal compared with the observed Cepheid light curve.  The failure is therefore not only an amplitude deficit, but also a morphology deficit.
    
    \item Models with \(\alfat=0.0975\) and \(\alfat=0.100\) are treated as diagnostic rather than accepted because they did not satisfy the adopted quality-control criteria.

    \item The main scientific contribution is a reproducible period-calibrated and morphology-tested \MESA-\RSP{} model sequence showing that the accepted models remain deficient in harmonic content, rise-time asymmetry, and observed-band amplitude.  A complete observed-light-curve solution for \dcep{} therefore remains a target for further modeling.
\end{enumerate}

\section*{Data and Code Availability}
\label{sec:data_code_availability}

The numerical products underlying this analysis include the final inlists, reduced light-curve tables, Fourier diagnostics, plotting scripts, and figure files for the accepted and diagnostic model sequence. The principal scripts used for this stage are \code{make_phase_template_alfat0p095_500.py}, \code{make_mist_bc_synthetic_V_curve_alfat0p095.py}, \code{measure_lightcurve_asymmetry.py}, and \code{make_final_morphology_comparison.py}.

\section*{Acknowledgments}
\label{sec:acknowledgments}

The authors acknowledge the developers of \MESA{} and the broader stellar pulsation community whose tools and methods made this work possible.


\bibliographystyle{unsrtnat}
\bibliography{references}

\end{document}